\documentclass[electronic]{vgtc}                  
\ifpdf
  \pdfoutput=1\relax                   
  \usepackage{graphicx}                
  \DeclareGraphicsExtensions{.pdf,.png,.jpg,.jpeg} 
\else
  \usepackage{graphicx}                
  \DeclareGraphicsExtensions{.eps}     
\fi%

\graphicspath{{figures/}{pictures/}{images/}{./}} 

\usepackage{microtype}                 
\PassOptionsToPackage{warn}{textcomp}  
\usepackage{textcomp}                  
\usepackage{mathptmx}                  
\usepackage{times}                     
\usepackage{cite}                      
\usepackage{tabu}                      
\usepackage{booktabs}   
\newcommand{\nop}[1]{}

\usepackage[normalem]{ulem}
\useunder{\uline}{\ul}{}

\newcommand{\etal}{\emph{et~al. }}
\newcommand{\eg}{\emph{e.g., }}
\newcommand{\ie}{\emph{i.e., }}
\usepackage{todonotes}

\onlineid{0}

\vgtccategory{Research}

\vgtcinsertpkg

\title{SAGE: Intrusion Alert-driven Attack Graph Extractor}

\author{Azqa Nadeem\thanks{e-mail: azqa.nadeem@tudelft.nl}\\ %
        \scriptsize Delft University of Technology %
\and Sicco Verwer\thanks{e-mail: s.e.verwer@tudelft.nl}\\ %
     \scriptsize Delft University of Technology %
\and Shanchieh~Jay~Yang\thanks{e-mail: Jay.Yang@rit.edu}\\ %
     \scriptsize Rochester Institute of Technology}

\abstract{Attack graphs (AG) are used to assess pathways availed by cyber adversaries to penetrate a network. State-of-the-art approaches for AG generation focus mostly on deriving dependencies between system vulnerabilities based on network scans and expert knowledge. In real-world operations however, it is costly and ineffective to rely on constant vulnerability scanning and expert-crafted AGs.
We propose to automatically learn AGs based on actions observed through intrusion alerts, without prior expert knowledge. Specifically, we develop an unsupervised sequence learning system, SAGE, that leverages the temporal and probabilistic dependence between alerts in a suffix-based probabilistic deterministic finite automaton (S-PDFA) -- a model that accentuates infrequent severe alerts and summarizes paths leading to them. AGs are then derived from the S-PDFA on a per-objective, per-victim basis.
Tested with intrusion alerts collected through Collegiate Penetration Testing Competition, SAGE compresses over 330k alerts into 93 AGs. These AGs reflect the strategies used by the participating teams.  The AGs are succinct, interpretable, and capture behavioral dynamics, \eg that attackers will often follow shorter paths to re-exploit objectives.%
} 

\CCScatlist{
    \CCScatTwelve{Security and privacy}{Intrusion/anomaly detection and malware mitigation}{Intrusion detection systems}{};
  \CCScatTwelve{Human-centered computing}{Visu\-al\-iza\-tion}{Visualization application domains}{Visual analytics};
  \CCScatTwelve{Computing methodologies}{Machine learning}{Learning paradigms}{Unsupervised learning};
}

\begin{document}

\firstsection{Introduction}\label{sec:intro}

\maketitle

Security Operation Centers (SOCs) typically receive thousands of intrusion alerts on a daily basis\footnote{https://blog.paloaltonetworks.com/2020/09/secops-analyst-burnout/}. While alert correlation techniques help to reduce alerts from intrusion detection systems (IDS)~\cite{alserhani2016alert,salah2013model,sadoddin2006alert}, they do not show the attack progression and attacker strategies, \ie they show \textit{what} the attackers did, but do not provide insight into \textit{how} the infrastructure was exploited. 

Attack graphs (AG) are models of attacker strategies that have been widely used for visual analytics~\cite{chu2010visualizing,angelini2015percival} and network hardening~\cite{jha2002two, kaynar2016taxonomy}. Existing approaches to generate AGs are expensive due to their expert-driven nature --- they utilize a significant amount of prior knowledge~\cite{ning2002constructing,ning2004building,alserhani2016alert} and published vulnerability reports~\cite{ou2005mulval,roschke2011new,gao2018exploring,hu2020attack}. In real-world operations however, it is costly and ineffective to rely on constant vulnerability scanning and expert-crafted AGs. Meanwhile, SOCs often possess large volumes of intrusion alerts from prior security incidents. We show, for the first time, that these alerts can be used as a basis to generate AGs. 

\begin{figure*}[t]
    \centering
    \includegraphics[width=0.75\linewidth]{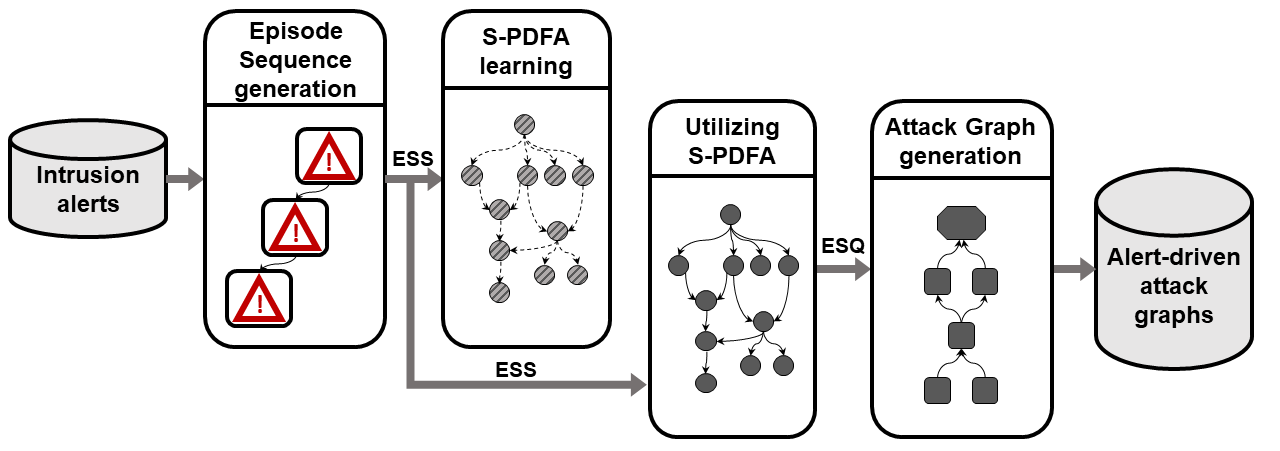}
    \caption{SAGE workflow: Intrusion alerts go in, attack graphs come out. An S-PDFA is learned in the first phase, and the model is utilized in the second phase to extract \textit{alert-driven attack graphs}.}
    \label{fig:workflow}
\end{figure*}

In this paper, we propose SAGE --- Intru\underline{S}ion alert-driven \underline{A}ttack \underline{G}raph \underline{E}xtractor. SAGE leverages sequence learning to mine patterns from intrusion alerts, models them using an automaton, and represents them in the form of an attack graph. The two core phases of SAGE are shown in Figure \ref{fig:workflow}. A tool such as SAGE can have important implications for the training and evaluation of defensive controls. SAGE augments existing IDSs by compressing several thousands of alerts into a handful of AGs. SOC analysts can triage alerts by visualizing and filtering AGs of interest. 

Class imbalance presents a major challenge for machine learning (ML) based attacker strategy identification --- severe alerts are scarce, and low-severity alerts (produced by network scans) are prevalent, which are not very valuable for analysts~\cite{valeur2004comprehensive}. While most ML solutions discard infrequent patterns, we propose a suffix-based probabilistic deterministic finite automaton (S-PDFA) --- a model that accentuates infrequent severe alerts, without discarding low-severity alerts. The S-PDFA summarizes attack paths leading to severe attack stages. It can differentiate between alerts that have identical signatures but different contexts, \eg scanning at the start, and scanning midway through an attack are treated differently because the former indicates reconnaissance and the latter indicates attack progression. AGs are then extracted from the S-PDFA on a per-objective, per-victim basis. A vertex in an AG represents a group of alerts generated by an attacker action, and an edge captures the temporal relationship between actions (as determined by the S-PDFA), showing attack progression. These graphs not only enable forensic analysis of prior security incidents, they also unlock a new means to derive intelligence regarding attacker strategies without having to investigate thousands of intrusion alerts.

We demonstrate the effectiveness of SAGE on distributed multi-stage attack scenarios, \ie where multi-member teams progress through various attack stages in order to compromise numerous targets. Penetration testing competitions provide an ideal setting to study such attacks. To this end, we use open-source intrusion alerts collected through Collegiate Penetration Testing Competition (CPTC)\footnote{https://www.globalcptc.org/}. 93 AGs are generated from $\sim$330k alerts. The AGs reflect the actual pathways taken by the penetration testers, even with an imperfect IDS. They are succinct, and effective in highlighting strategic differences between participating teams. They reveal behavioral dynamics, \eg that attackers often follow shorter paths to re-exploit an objective after they have discovered a longer one. We also show how to rank attackers based on the uniqueness and severity of their actions. Thus, our contributions are: 
\begin{enumerate}
    \item We develop SAGE, a tool that automatically generates succinct high-severity attack graphs from intrusion alerts, without prior knowledge about vulnerabilities or network topology. 
    \item We apply SAGE on alerts from a penetration testing competition. The AGs are effective in attacker strategy comparison.
\end{enumerate}

\section{Related work}\label{sec:litrev}

Attack graphs (AG) are a frequent area of research in the VizSec community. Kaynar \etal~\cite{kaynar2016taxonomy} proposed a taxonomy of the existing expert-driven AG generation approaches in the network security domain. These approaches generally utilize a knowledge base, making them unsuitable for zero-day vulnerabilities. Specifically, MulVAL~\cite{ou2005mulval} is an attack graph generator that has been widely used as a foundation for other works~\cite{roschke2011new,hu2020attack}, which takes the network topology and vulnerability information as input. Other techniques focus on path reachability~\cite{williams2008garnet,chu2010visualizing} and complexity reduction of expert-driven AGs~\cite{ingols2009modeling,homer2008improving}, instead of exploring additional data sources for AG construction. 
In addition, \textit{Process mining} (PM) has been used to visualize alert datasets~\cite{de2018process,chen2020distributed} without actually extracting AGs. \textit{Hidden markov models} (HMM) have been used to build alert forecasting systems~\cite{ghafir2019hidden}, and \textit{markov chains} have been used to build alert correlation systems~\cite{fredj2015realistic}. Specifically, Moskal \etal~\cite{moskal2018extracting} have used markov chains to model attacker strategies from intrusion alerts in the form of sequences. They use Jenson-Shannon divergence to measure similarity between such sequences. These approaches do not construct AGs and have several shortcomings if used to this aim: PM uses alert signatures as identifiers, which makes it impossible to differentiate alerts with identical signatures but different contexts, \ie those that lead to different paths. Markov chains have a similar weakness. HMMs do model context, but they are considerably difficult to interpret due to their non-deterministic nature. 
In this paper, we borrow initial ideas from~\cite{moskal2018extracting} and leverage the temporal and probabilistic dependence between alerts to generate alert-driven AGs. The probabilistic deterministic finite automaton (S-PDFA) that SAGE uses has more expressive power than markov chains, \ie it does model context, and is easier to interpret. 
We show how to construct objective-oriented AGs that visualize large volumes of alerts without high cognitive load. To the best of our knowledge, SAGE is the first successful approach for this challenging problem.
\section{Alert-driven Attack Graphs}\label{sec:method}
SAGE (Intru\underline{S}ion alert-driven \underline{A}ttack \underline{G}raph \underline{E}xtractor)\footnote{\url{https://github.com/tudelft-cda-lab/SAGE}} takes raw intrusion alerts as input, and transforms them into aggregated sequences that are used to learn a model summarizing attack paths in the data. Attack graphs (AG) are extracted from this model on a per-objective, per-victim basis (see Figure \ref{fig:workflow}).  
The AGs are succinct and interpretable as they compress large volumes of alerts in order to show how an attack transpired. They also provide an effective means for attacker strategy comparison. SAGE is agnostic to host and network properties. It is released as a docker container for cross-platform support.

The first step towards building AGs is to arrange intrusion alerts in sequences that characterize an attacker strategy. 
An IDS alert contains 
the attacker and victim IP addresses, the targeted service $tServ$ derived from destination port\footnote{Derived from open-source Port$\rightarrow$Service mapping from IANA.}, 
and the attack stage $mcat$ derived from the existing Action-Intent framework by Moskal \etal~\cite{moskal2020framework} (see appendix), based on MITRE ATT\&CK~\cite{strom2018mitre}. 
%
Raw intrusion alerts are often noisy and contain duplicates. Thus, cleaning and aggregating them is necessary. 
We aggregate alerts into groups, such that 
they likely belong to the same attacker action. In literature, such an aggregation is called a hyper-alert or an attack episode. Grouping alerts that appear in bursts is a common way to construct episodes. We use the method in~\cite{moskal2018extracting} to aggregate alerts into \textit{attack episodes}, and assume that the episodes closely characterize attacker actions.
For an attack stage $mcat$, an episode is defined as $\langle st, et, mcat, mServ \rangle$, where $st$ and $et$ are start/end times, and $mServ$ is the most frequently targeted service during the episode. 
We create time-sorted \textit{episode sequences} (ES) for each (attacker,victim) combination, and partition the ES whenever a low-severity episode follows a high-severity one, signaling the start of a new attack attempt (see $mcat$ $\rightarrow$ severity mapping in appendix). These are called the \textit{episode sub-sequences} (ESS).

We propose a \textit{suffix-based probabilistic deterministic finite automaton} (S-PDFA), which is a suffix-variant of the probabilistic deterministic finite automaton~\cite{verwer2014pautomac}. Instead of predicting the future, the S-PDFA can be used to predict the past. Since the high-severity $mcat$'s are at the end of episode sub-sequences, we specifically learn a suffix model to determine which episodes eventually lead to high-severity attack stages. We provide all the ESS's from CPTC-2018 to the Flexfringe automaton learning framework~\cite{verwer2017flexfringe}. Flexfringe uses univariate \textit{symbol} sequences comprised of $\langle mcat, mServ\rangle$ to learn the S-PDFA (see appendix). The model summarizes attack paths in the dataset and clusters them based on behavioral similarity. It also brings infrequent high-severity actions into the spotlight, without discarding low-severity ones. This is tricky because while most clustering approaches discard infrequent patterns, the S-PDFA salvages them by setting appropriate parameters in Flexfringe (see Section \ref{sec:expsetup}), which also results in an interpretable model.

\begin{figure}[h]
    \centering
    \includegraphics[width=0.85\linewidth]{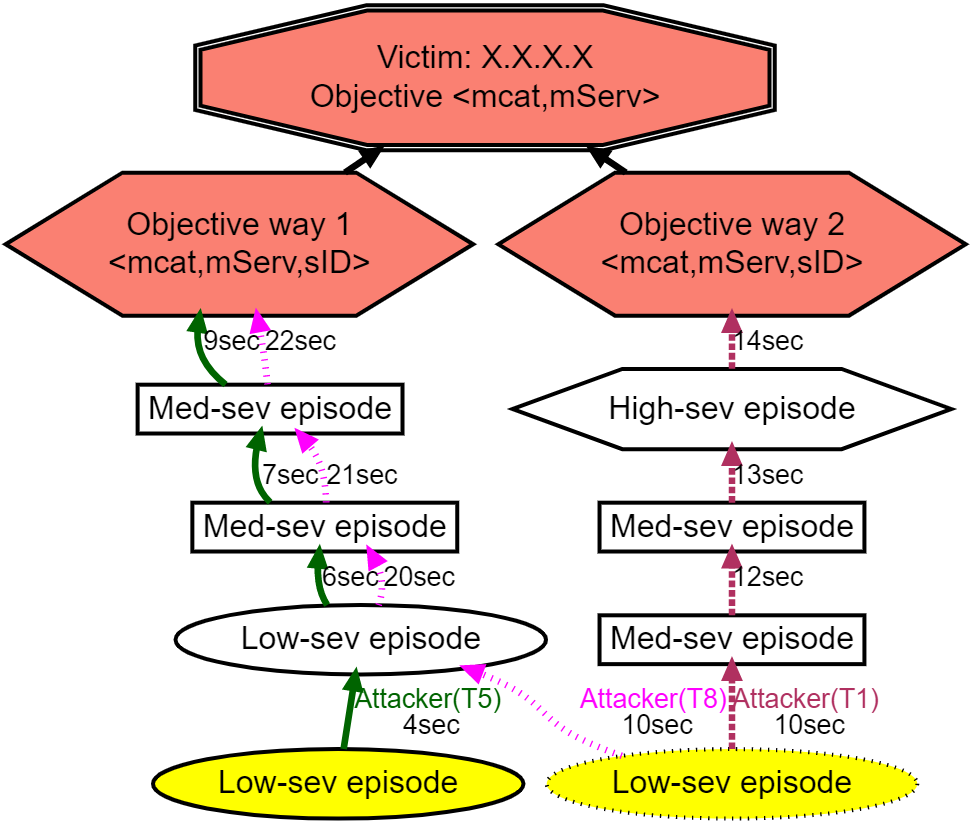}
    \caption{Notional alert-driven AG showing paths towards an objective. \emph{Vertex labels} are $\langle mcat, mServ, sID\rangle$, \ie attack stage, targeted service and state identifier. Low-severity actions are \textit{ovals}, medium-severity are \textit{boxes}, high-severity are \textit{hexagons}.  The first action in a path is \textit{yellow}, while the objective-variants are \textit{red}. Actions that occur too infrequently for Flexfringe are \textit{dotted}. Edge label shows time since first alert. Edge style shows team: T1 (\textit{Dashed}), T5 (\textit{Solid}), T8 (\textit{Dotted}).}
    \label{fig:exampleAG}
\end{figure}

The states in an S-PDFA can be considered as milestones achieved by attackers,  
providing contextual meaning to the episodes' attack stages. Prior work by Lin \etal~\cite{lin2018moha} has utilized this context to cluster similar car-following behaviors. We follow the same idea and convert episode sequences into state sequences (ESQ): we replay each ESS through the S-PDFA and augment it with state identifiers ($sID$), resulting in its corresponding ESQ.
Finally, the ESQs are transformed into a graph (AG) via Graphviz (see Figure~\ref{fig:exampleAG}). These graphs are generated on a per-objective ($obj$), per-victim ($vic$) basis. An $obj$ is defined as $\langle mcat,mServ,sID\rangle$, \ie one of the high-severity attack stages from~\cite{moskal2020framework} (since they specify end-goals), the targeted service, and the state identifier. For an AG with the root vertex $\langle vic, obj\rangle$, only the ESQs concerning the victim $vic$, and containing an episode with $obj$ are included. 
If an $obj$ is achieved multiple times in an ESQ, each attempt is shown as an individual path in the graph. The S-PDFA may assign different $sID$'s to the same $\langle mcat, mServ\rangle$, corresponding to the different contextual means of obtaining the $obj$, each of which appears as a sub-objective in the graph. Also, all attackers that obtain $obj$ are shown in one graph to aid strategy comparison. 
Thus, an AG is a compressed representation of intrusion alerts related to $\langle vic, obj\rangle$. It shows how an attack transpired, including similarities between attacker strategies. 

\section{Experimental Setup}\label{sec:expsetup}
\textbf{Dataset.} We generate attack graphs for the open-source Collegiate Penetration Testing Competition dataset, \ie CPTC-2018~\cite{munaiah2019characterizing}. It contains Suricata alerts generated by different student teams tasked with compromising a fictitious network, \ie an automotive company. Each team has access to fixed-IP machines. Beyond the attackers' IP information, no ground truth is available regarding the attacker strategies and attack progression. Six teams (\ie T1, T2, T5, T7, T8, T9) produce 330,270 alerts. 
The competition lasted 9 hours. 

\noindent\textbf{Parameter selection.} SAGE has five parameters: we set $t = 1.0$ sec to discard repeated alerts~\cite{moskal2018extracting}, and window length $w=150$ sec to aggregate alerts into episodes. We set three Flexfringe parameters: $symbol\_count$, $state\_count$ and $sink\_count$, all set to 5. These parameters are selected based on the properties of the dataset, primarily dependent on the frequency of severe alerts.
The experiments are executed on Intel Xeon W-2123 quad-core processor, 32 GB RAM.

\noindent\textbf{S-PDFA model quality.} Quantifying S-PDFA model quality is a difficult problem~\cite{parekh2001learning,de2010grammatical}. A common option is to measure its prediction power using \textit{Perplexity}~\cite{verwer2014pautomac,balle2017results}. 
Compared with suffix trees and markov chains, our S-PDFA achieves the best perplexity, showing its ability to capture patterns in the sequences (see appendix).

\begin{table}[t]
\centering
\caption{Workload reduction in the CPTC-2018 dataset.}
\resizebox{0.95\columnwidth}{!}{
\begin{tabular}{|c|c|c|c|c|c|c|}
\hline
\textbf{} & \textbf{\begin{tabular}[c]{@{}c@{}}Alerts \\ (raw)\end{tabular}} & \textbf{\begin{tabular}[c]{@{}c@{}}Alerts \\ (filtered)\end{tabular}} & \textbf{Episodes} & \textbf{\begin{tabular}[c]{@{}c@{}}ES/ \\ ESQ\end{tabular}} & \textbf{ESS} & \textbf{AGs}  \\ \hline
\textbf{T1} & 81373 & 26651 &  655 &  103 & 108 &  53\\ \hline
\textbf{T2} & 42474 & 4922 &  609 & 86 & 92 & 7\\ \hline
\textbf{T5} & 52550 & 11918 &   622 & 69 & 74 & 51\\ \hline
\textbf{T7} & 47101 & 8517 & 576& 63 & 73 & 23\\ \hline
\textbf{T8} & 55170 &   9037 &  439& 67 & 79 & 33\\ \hline 
\textbf{T9} & 51602 &   10081 &  1042& 69& 110 & 30  \\ \hline 
\end{tabular}}
\label{tab:data-summary-teamlvl}
\end{table}

\section{Results and Discussion}\label{sec:results}
SAGE compresses 330,270 alerts into 93 attack graphs (AG), and discovers 70 objectives that are obtained by targeting 19 victim hosts. This leads to a considerable workload reduction. Table \ref{tab:data-summary-teamlvl} shows this reduction on a per-team basis. Note that multiple teams can share one AG if they all obtain that objective. Furthermore, the AGs are succinct and effective in highlighting differences between attacker strategies. We evaluate the complexity of the AGs using the model simplicity metric given in~\cite{de2018process}, \ie $Simplicity(AG) = \frac{|V|}{|E|}$, where $|V|$ and $|E|$ are the number of vertices and edges, respectively. The AGs have an average simplicity of 0.81, with 21.7 vertices on average, where AGs with more than 30 vertices are considered as complex~\cite{de2018process}. Each AG represents about 500 alerts, on average.

\begin{figure}[t]
    \centering
    \includegraphics[width=\linewidth]{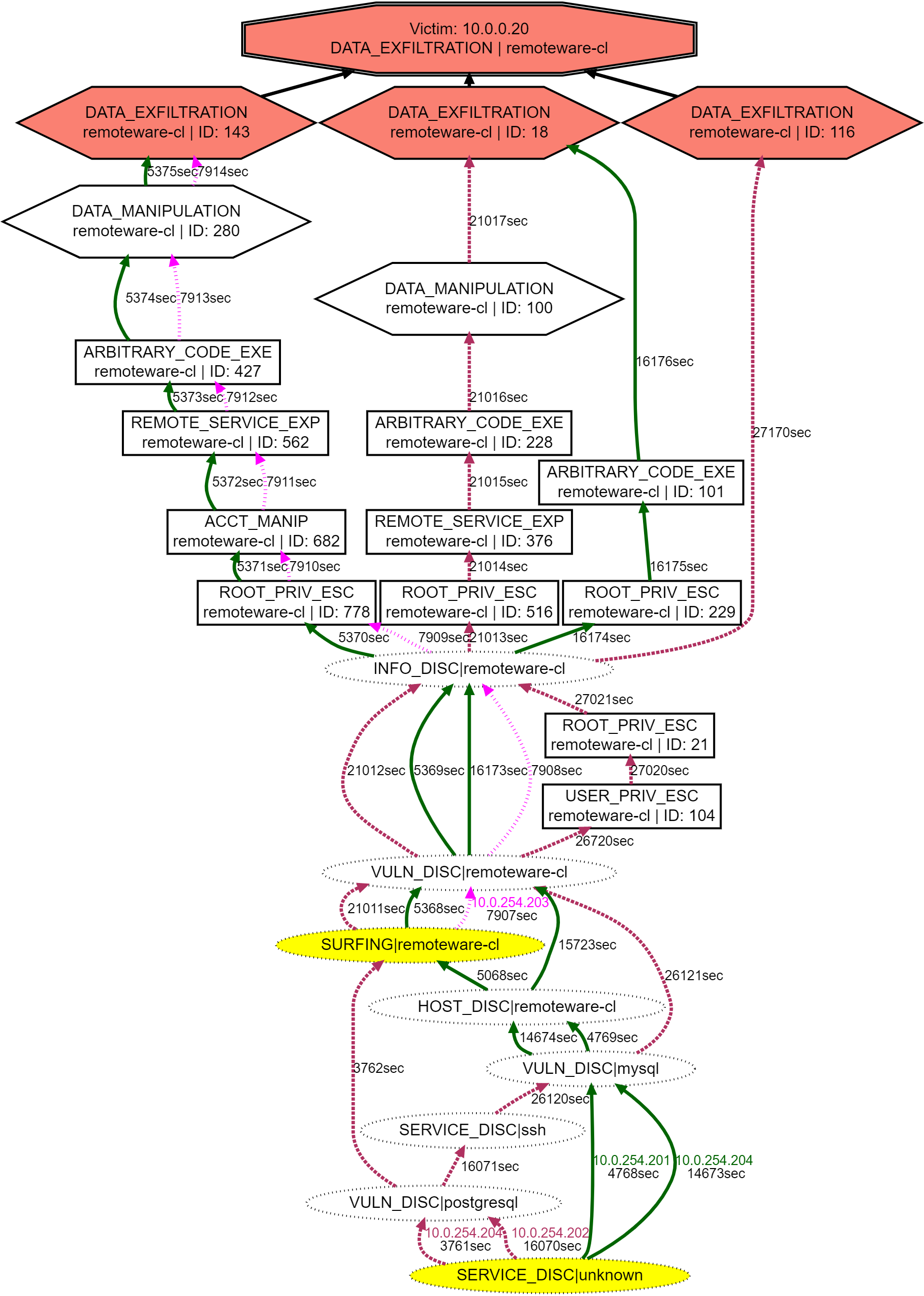}
    \caption{Data exfiltration/remoteware by T1, T5, and T8. T1 and T5 make two attempts, while T8 makes a single attempt. The S-PDFA discovers three contextual ways of exfiltrating data from this victim.}
    \label{fig:cptc-ag-1}
\end{figure}

\textbf{1. AGs show attack pathways:} The AGs provide insight into attacker strategies. Figure \ref{fig:cptc-ag-1} shows that three teams (T1, T5, T8) use remoteware-cl to exfiltrate data from 10.0.0.20 (the absence of other teams means they were unable to obtain this objective). The teams self-reported that they had found a chatting application on this host that contained credentials, which were exfiltrated using a combination of privilege escalation and arbitrary code execution. 
T1 finds two distinct paths to complete this objective, first after 5.8 hours and then again after 7.5 hours since the start of the competition. T5 also finds two paths, but considerably earlier in the competition. 
The S-PDFA identifies three contextually distinct exfiltration states based on the differences in the paths that lead up to the objective. For example, \texttt{$\langle$data\_exfiltration, remoteware-cl, 116$\rangle$} can be reached with much fewer steps compared to the others, and it also happens much later in the competition, implicitly capturing attackers' increasing experience. Moreover, in cases where multiple attack attempts are made, the subsequent attempt is shorter than the first in 84.5\% of the cases, providing evidence for SAGE's ability to capture behavioral dynamics. 

\begin{figure}[t]
    \centering
    \includegraphics[width=\linewidth]{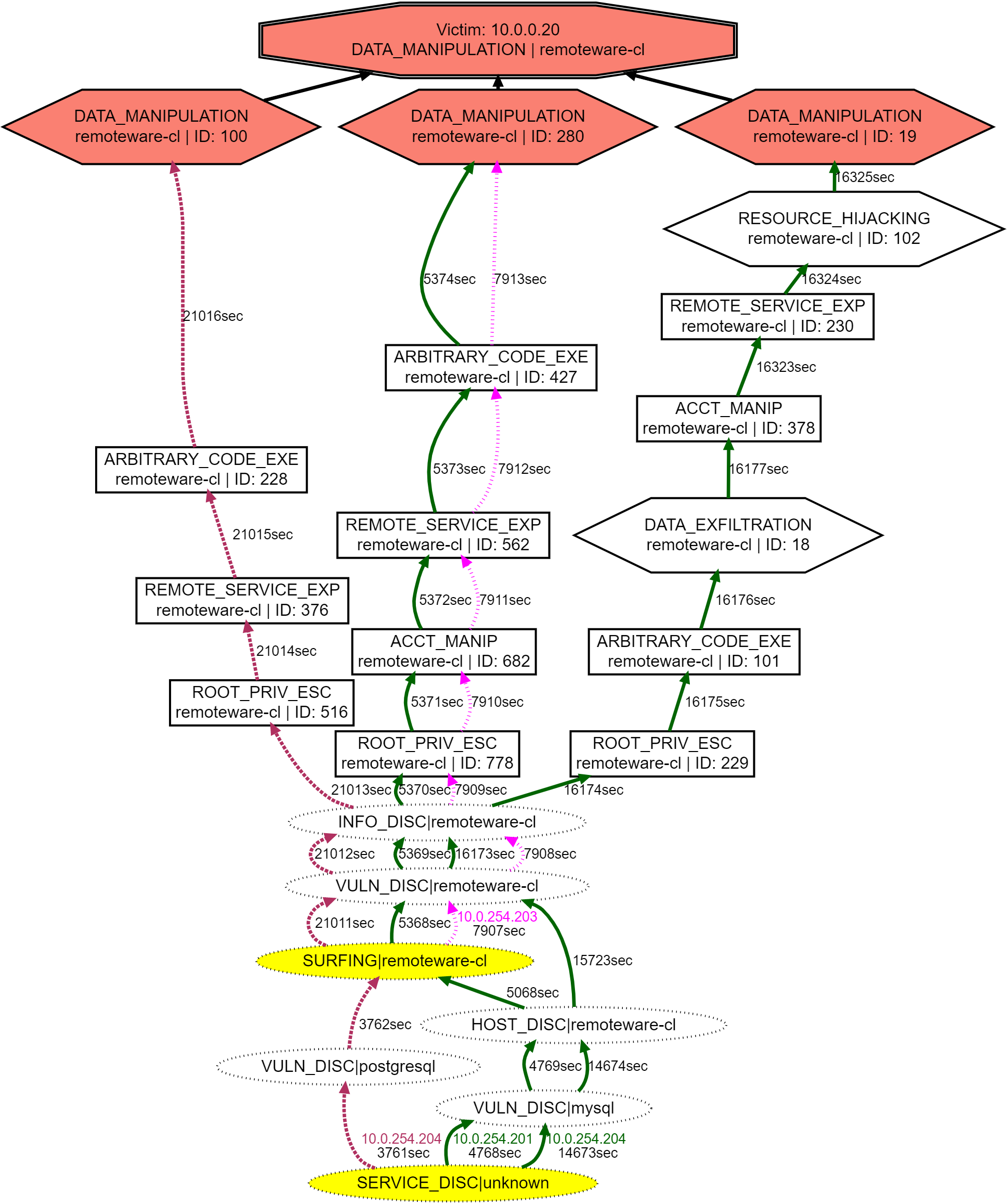}
    \caption{Data manipulation/remoteware is a sub-graph of Figure \ref{fig:cptc-ag-1}.}
    \label{fig:append-cptc-ag-1}
\end{figure}

\textbf{2. AGs show strategic differences:} Interestingly, an AG of data manipulation (Figure \ref{fig:append-cptc-ag-1}) over the same victim and service from Figure \ref{fig:cptc-ag-1} are partial sub-graphs of each other, due to overlap in paths that attain both objectives. There are three variants of data manipulation, of which two are also present in the exfiltration AG, \ie \texttt{$\langle$data\_manipulation, remoteware-cl, 100$\rangle$} and \texttt{$\langle$data\_manipulation, remoteware-cl, 280$\rangle$}. T5 finds an additional path to reach \texttt{$\langle$data\_manipulation, remoteware-cl, 19$\rangle$} after it has reached the objective \texttt{$\langle$data\_exfiltration, remoteware-cl, 18$\rangle$} from the previous AG. This actionable intelligence can be used to disrupt the cyber kill-chain~\cite{hutchins2011intelligence}. Additionally, the AG shows differences in attacker strategies, \eg T5 and T8 perform account manipulation while T1 does not; and resource hijacking is a step in one of T5's paths but not in the other. It also shows that T1 has found the shortest path to perform data manipulation on the victim using remoteware-cl.

\textbf{3. AGs allow attacker performance evaluation:} Each vertex in the attack graphs signifies a new milestone achieved by the teams. We argue that the fraction of unique milestones, \ie $\langle mcat, mServ, sID\rangle$, discovered by a team provides a metric for its performance. A medium-severity attack stage forms a stepping-stone towards a high-severity attack stage. Hence, high-severity vertices are twice as important as medium-severity vertices, \ie $\frac{(2*sev) + (1*med)}{3}$, where $sev$ and $med$ are the number of high- and medium-severity milestones discovered by a team, respectively. 
Table \ref{tab:eval-of-teams} shows the evaluation of all six teams based on the 93 AGs, ranked according to their score. It shows the number of unique high- and medium-severity vertices discovered by the teams during the competition. 
T5 scores the highest points, while T2 scores the lowest points. T1 comes in second, solely because they discover the highest number of medium-severity vertices compared to any other team. Overall, this metric provides a simple way to rank attackers based on the uniqueness and severity of the alerts they raise. 

\begin{table}[t]
\centering
\caption{Attacker evaluation based on the fraction of unique vertices discovered during CPTC-2018 (\textit{rank = score}).}
\resizebox{\columnwidth}{!}{
\begin{tabular}{|c|c|c|c|}
\hline
\textbf{Team} &
  \textbf{\begin{tabular}[c]{@{}c@{}}Severe vertices\\ (out of 70)\end{tabular}} &
  \textbf{\begin{tabular}[c]{@{}c@{}}Medium vertices\\ (out of 148)\end{tabular}} &
  \textbf{\begin{tabular}[c]{@{}c@{}}Weighted average\\ percentage\end{tabular}} \\ \hline
\textbf{T5} & 28 (40\%) & 40 (27\%) & 35.67 \\ \hline
\textbf{T1} & 18 (26\%) & 62 (42\%) & 31.33 \\ \hline
\textbf{T9} & 23 (33\%) & 36 (24\%) & 30.0  \\ \hline
\textbf{T7} & 22 (31\%) & 26 (18\%) & 26.67 \\ \hline
\textbf{T8} & 15 (21\%) & 32 (22\%) & 21.33 \\ \hline
\textbf{T2} & 3 (4\%)   & 8 (5\%)   & 4.33  \\ \hline
\end{tabular}}
\label{tab:eval-of-teams}
\end{table}

\subsection{Discussion: Visual Analytics enabled by SAGE}
Utilizing observables such as intrusion alerts to obtain intelligence regarding attacker strategies will noticeably benefit SOC analysts. Visualizing such strategies in a way that communicates the correct message to SOC analysts is another important challenge. SAGE is one of the first attempts toward addressing these challenges: SAGE extracts targeted (objective-oriented) attack graphs (AG) from intrusion alerts without prior knowledge. In doing so, SAGE opens up numerous research opportunities for the VizSec community. 
\begin{itemize}
    \item Although SAGE significantly reduces the volume of alerts to analyze, it still ends up with several AGs. The prioritization of AGs in general, and attack paths in particular, remains an open problem. A query mechanism to filter and replay specific parts of an attack, \eg vaguely similar to PERCIVAL~\cite{angelini2015percival}, will be highly useful. 
    \item SAGE can be used to improve IDS rules. It utilizes most of the alerts, which are aggregated for visual analytics. Executing test attacks for which no corresponding paths can be found in the resulting AGs hint towards missing or faulty IDS rules.  
    \item Comparative visual analytics for AGs is another open challenge. For a given victim that is attacked by different attackers at different times, highlighting the attack progression similarity for visual comparison is an interesting direction.
    \item In forensic analysis, alert-driven AGs can be used to point towards specific victim machines for which additional evidence is required. This evidence can be used to corroborate the success of critical milestones for an investigated attack. For example, an AG containing data exfiltration step highlights the need for investigating other data sources to check whether data were indeed exfiltrated. 
\end{itemize}

\section{Conclusions and future work}\label{sec:conclusion}
In this paper, we propose SAGE, a novel unsupervised sequence learning tool that generates succinct attack graphs (AG) directly from raw intrusion alerts, without a priori knowledge. SAGE is capable of representing several thousands of alerts in just a handful of AGs, which is beneficial for alert triaging and visual analytics. SAGE models the temporal and probabilistic dependence between alerts in a suffix-based probabilistic deterministic finite automaton (S-PDFA). The S-PDFA brings infrequent severe alerts into the spotlight and summarizes paths leading to them. AGs are then extracted from the S-PDFA on a per-objective, per-victim basis.
SAGE generates 93 AGs from $\sim$330k alerts collected through Collegiate Penetration Testing Competition with six attacker teams. The AGs provide a clear picture of the attack progression, and show strategic differences between attackers, \eg they show that attackers often follow shorter paths to re-exploit an objective. They are also used to rank interesting attackers based on the severity of the alerts they raise.

Future work will focus on applying SAGE to additional datasets, adding interactive capabilities to alert-driven AGs, evaluating AGs with security analysts, and leveraging readily-available domain knowledge to map AG vertices to high-level attacker actions.

\bibliographystyle{abbrv-doi-hyperref-narrow}
\bibliography{template}

\appendix
\section{Appendix}

\subsection{Attack stages}

Table \ref{tab:mas-meaning} provides the Action-Intent mapping that is used by SAGE to map default alert signatures to attack stages.

\begin{table}[ht]
\centering
\caption{Attack stages and their severity from Moksal \etal}
\scalebox{0.7}{
\begin{tabular}{|l|l|l|}
\hline
\textbf{Acronym} & \textbf{Attack stage} & \textbf{Severity} \\ \hline
SURFING  & Surfing & Low\\ \hline
HOST\_DISC  & Host Discovery & Low\\ \hline
SERVICE\_DISC  & Service Discovery & Low\\ \hline
VULN\_DISC  & Vulnerability Discovery & Low\\ \hline
INFO\_DISC  & Information Discovery & Low\\ \hline
USER\_PRIV\_ESC  & User Privilege Escalation & Med\\ \hline
ROOT\_PRIV\_ESC  & Root Privilege escalation & Med\\ \hline
BRUTE\_FORCE\_CREDS  & Brute force Credentials & Med\\ \hline
ACCT\_MANIP  & Account Manipulation & Med\\ \hline
PUBLIC\_APP\_EXP  & Public Application Exploitation & Med\\ \hline
REMOTE\_SERVICE\_EXP  & Remote Service Exploitation & Med\\ \hline
COMMAND\_AND\_CONTROL  & Command and Control & Med\\ \hline
LATERAL\_MOVEMENT  & Lateral movement & Med\\ \hline
ARBITRARY\_CODE\_EXE  & Arbitrary code execution & Med\\ \hline
PRIV\_ESC  & Privilege escalation & Med\\ \hline
NETWORK\_DOS &  Network Denial of Service & High\\ \hline
RESOURCE\_HIJACKING & Resource hijacking & High\\ \hline
DATA\_MANIPULATION  & Data manipulation & High\\ \hline
DATA\_EXFILTRATION  & Data exfiltration & High\\ \hline
DATA\_DELIVERY  & Data delivery & High\\ \hline
DATA\_DESTRUCTION  & Data destruction & High\\ \hline
\end{tabular}}
\label{tab:mas-meaning}
\end{table}

\subsection{S-PDFA for CPTC-2018}
\begin{figure}[t]
    \centering
    \includegraphics[width=\linewidth]{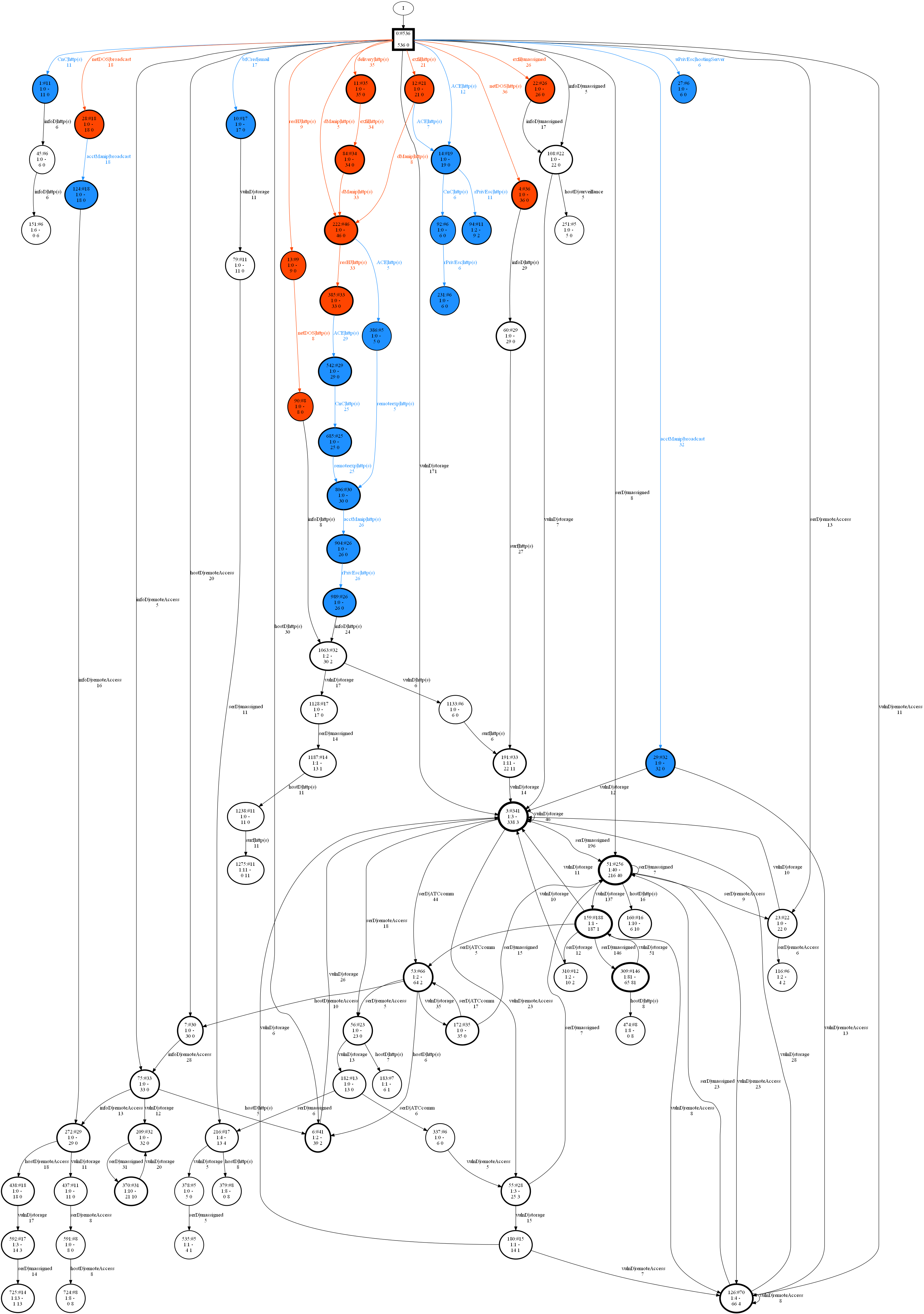}
    \caption{The S-PDFA for 6 teams in CPTC-2018. State color denotes severity: red = high, blue = medium, white = low.}
    \label{fig:dfa-model}
\end{figure}

Figure \ref{fig:dfa-model} shows the S-PDFA learned for the entire CPTC-2018. 

\subsection{S-PDFA model quality}
Model quality is often quantified using a model selection criterion, measuring a trade-off between model size and fit. 
Perplexity is defined as $2^{-\frac{1}{N}\sum_{i=1}^{N}log_2 P(x_i)}$ where $N$ is the number of traces and $P(x_i)$ returns the probability of the $x_i$ trace. \textit{The lower the value, the better the model fits with the data}. We compute perplexity for both the training data and an unseen test set using an 80-20 split. The former shows how well the model fits with the training data, and the latter shows how well the model captures patterns in the overall data. 
We compare the perplexity values against two suffix models: (a) suffix tree and (b) markov chains. Table \ref{tab:perplexity} shows the perplexity for each variant on both training and test data. It shows that a suffix tree provides the best fit with the training data, as expected. The S-PDFA is about twice as ``perplexed". On the test data, the S-PDFA gives the best perplexity value, demonstrating that the model accurately captures many of the patterns present in the data that are missed by the other models.

\begin{table}[h]
\centering
\caption{Perplexity of suffix models (bold = best value).}
\label{tab:perplexity}
\begin{tabular}{|l|c|c|c|c|c|}
\hline
\textbf{} & \multicolumn{1}{c|}{\textbf{Suffix tree}} & \multicolumn{1}{c|}{\textbf{Markov chains}} & \multicolumn{1}{c|}{\textbf{S-PDFA}} \\ \hline
\textbf{Training data} & \textit{\textbf{1265.4}} & 13659.6 & 2397.8 \\ \hline
\textbf{Holdout test set} & 13020.7 & 11617.8 & \textit{\textbf{9884.6}} \\ \hline
\end{tabular}
\end{table}

\end{document}